# Multiple criteria decision analysis with the SRF-II method to compare hypotheses of adaptive reuse for an iconic historical building

Francesca Abastante[a], Salvatore Corrente[b], Salvatore Greco[b,c], Isabella M. Lami[a], Beatrice Mecca[a*]

[a] *Department of Regional and Urban Studies and Planning (DIST), Politecnico di Torino, Viale Mattioli 39,10122 Torino, Italy.* francesca.abastante@polito.it *(F. Abastante),* isabella.lami@polito.it *(I.M. Lami),* beatrice.mecca@polito.it *(B. Mecca)*

[b] *Department of Economics and Business, University of Catania, Corso Italia, 55, 95129 Catania, Italy.* salvatore.corrente@unict.it *(S. Corrente),* salgreco@unict.it *(S. Greco)*

[c] *University of Portsmouth, Portsmouth Business School, Centre of Operations Research and Logistics (CORL), Richmond Building, Portland Street, Portsmouth PO1 3DE, United Kingdom*

Highlights:

- Conjunction of four Multi-Criteria Decision Analysis methods
- SRF– II method: improvement of SRF procedure for elicitation of weights of criteria
- Simulation of the decision-making process for the reuse of an iconic building
- Identification of the most appropriate design solution, among a set of alternatives

Abstract: The paper aims to show how Multiple Criteria Decision Aiding (MCDA) tools integrated in a well-articulated methodology can support the analyses of six hypotheses of adaptive reuse of an iconic historical building in Turin, Italy (called "Stock Exchange") to identify the preferred alternative of requalification. The debate around the requalification of the "Stock Exchange", conducted in the last two years, has been huge for several reasons: the building is perceived as an historical "monument" by the citizens; it shows extraordinary architectural and typological values with a high reputation at the national level; it involves public and private interests. In this context, interacting with experts involved in the ongoing discussion, we consider a recently proposed methodology based on the conjunction of four MCDA methods, namely: Multiple Criteria Hierarchy Process (MCHP), permitting to take into consideration structural relationships between criteria; ELECTRE III, permitting to judge if an alternative is at least as good as another taking into account reasons in favour and reasons against; the imprecise SRF method, supplying a easily understandable approach to collect information from the DM on the importance and the interaction of considered criteria; and Stochastic Multicriteria Acceptability Analysis (SMAA), permitting to take into account robustness concerns related to the range of variability of parameters of the model. An important point of our methodology is the consideration of three types of interaction effects between criteria: strengthening, weakening and antagonistic effects. With the aim of improving the comprehension of the information required to the DM to assess the weights of criteria, we propose a modification of the SRF methodology, called SFR-II, to increase the reliability of the decision aid procedure, which could constitute a significant advance for the same SRF method. As final result, our methodology provided robust recommendations in terms of probability of preference, indifference and incomparability between the project alternatives, at each level of the hierarchy of criteria. A discussion on the contribution that the MCDA methodology we adopted gives in the debate on the adaptive reuse of Stock Exchange as well as in analogous decision problems related to urban and territorial planning is also provided.

1. Introduction

The twentieth century faced a great rate of abandoned urban areas and buildings, whose transformation often constitutes a complex and problematic situation due to the presence of multiple objectives and different stakeholders that have to interact each other (e.g. owners, investors, public decision makers). In this context, from an architectural point of view, the adaptive reuse became a valuable approach for a new sustainable rebirth of the city (Dewiyana et al., 2016) encouraging the reuse of existing and abandoned sites and buildings, avoiding the waste of energy and materials caused by the new construction, preserving portion of urban landscape and offering new social and economic profits (Dewiyana et al., 2016).

In this framework, it is important to support decision analysis and decision-making by means of methodologies for which the information required as input and the recommendation supplied as output could be rigorous and accurate on one hand, but also as simple and understandable as possible on the other hand. This implies an appropriate "design" of the decision support procedure, in the perspective of the so-called "choice architecture for architecture choices" (Abastante et al. 2018). Accordingly, in this paper we apply a new multicriteria decision-making methodology (Corrente et al. 2017) to the architectural field, investigating the most suitable design solution for the reuse of an iconic historical building located in Turin (Italy): the Stock Exchange.

This particular case study is very promising since it belongs to the unused buildings asset, a current issue with an important dimension in Italy and more in general in Europe, which could be seen both as a problem and as a resource. Moreover, we had the opportunity to interface with some experts involved in the real ongoing decision-making process.

There is a complex debate behind the Stock Exchange, conducted in the last two years, since it stands out as a historical monument for the population, it presents architectural and typological values that need to be preserved and it involves financial and economic, public and private interests. In order to consider the richness of the debate, we decided to consider six possible transformations of the building, that have been analysed through a new multi-methodological framework to identify the preferred one with respect to the preference of the decision makers (DMs).

The methodology adopted, proposed by Corrente et al. (2017), is a conjunction of four Multiple Criteria Decision Aiding (MCDA) methods: Multiple Criteria Hierarchy Process (MCHP) (Corrente et al. 2012, 2013), ELECTRE III (Roy and Bouyssou 1993), imprecise Simos-Roy-Figueira (SRF) method (Figueira and Roy 2002; Corrente et al. 2017) and Stochastic Multicriteria Acceptability Analysis (SMAA) (Lahdelma et al. 1998). In this paper we suggest improving the SRF method introducing more natural questions for the DM, proposing the new "SRF-II method".

The choice of this decision support approach is related to several reasons: the management of a large number of criteria through the MCHP that takes into account the hierarchical structure of criteria on which the alternatives are evaluated; the consideration of three types of interaction effects between criteria (strengthening, weakening and antagonistic effects), through the ELECTRE III method; the manage imprecise preference information provided by the DMs, through the imprecise SRF-II method. According to the latter, mention has to be made to the innovation that we are introducing, proposing to consider a "zero criterion", with the aim of facilitating the DM in eliciting the weights of criteria. The improvement we are proposing has also a specific interest in the MCDA field, because it permits to collect a more homogenous information composed of only numbers of blank cards between levels of criteria with equal importance, avoiding an "ethereal" ratio of weights between the most and the least important criterion. Finally, the SMAA methodology provides robust recommendations taking into account the presence of a plurality of vectors of parameters (e.g. weights of criteria) compatible with the preferential information supplied by the DMs. These recommendations are expressed in terms of frequency in a certain number of computational simulations of preference, indifference and incomparability between the project alternatives, at each level of the hierarchy.

The paper is organised as follows: section 2 provides a description of the methodological and theoretical framework; section 3 describes the case study and the six alternatives of project, while section 4 illustrates the MCDA application and section 5 provides the results obtained. Finally, conclusions and future developments are provided in section 6.

2. Methodological framework

In MCDA (Greco et al., 2016) a set of alternatives $A = \{a, b, ...\}$ has to be evaluated on a set of aspects $G = \{g_1, ..., g_m\}$, technically called criteria, to deal with a choice, ranking or sorting problem. In this paper we are interested in ranking problems in which alternatives have to be partially or totally ordered from the best to the worst with the possibility of some ex-aequo among them.

Looking at the performances of the alternatives on the considered criteria, the only objective information that can be gathered is the dominance relation $D$ where $aDb$ if $a$ is at least as good as $b$ for all criteria and $a$ is better than $b$ for at least one of them. The objectivity of this relation is counterbalanced by its poverty since in comparing a pair of alternatives $a$ and $b$, it happens quite often that $a$ is better than $b$ on some criteria and $b$ is better than $a$ on some others, so that neither $aDb$ nor $bDa$. For such a reason, when comparing pairs of alternatives $a$ and $b$, the preferences with respect to considered criteria of $a$ over $b$ have to be to be aggregated and compared with the analogous preferences of $b$ over $a$ to get an overall comparison in terms of an outranking relation $S$ such that for each pair of alternatives $a$ and $b$ $aSb$ means that "$a$ is at least as good as $b$". On the basis of the comprehensive preference comparisons of all pairs of alternatives, using some appropriate procedure, a final recommendation on the considered problem can be defined (Roy 1990). Several MCDA methods have been introduced to this aim and, in this paper, we decided to apply the robust and hierarchical ELECTRE III method (Corrente et al. 2017), being the conjunction of four methods which will be presented in the next sections.

2.1 Multiple Criteria Hierarchy Process

In almost all real-world problems, the evaluation criteria are not at the same level but they are structured in a hierarchical way. It is therefore possible to define a root criterion (being the objective of the decision problem, corresponding to an overall evaluation of considered alternatives), some macro-criteria and other criteria descending from them.  For example, taking into account a decision problem related to the adaptive reuse of an historical building as is the case for this paper, one can imagine macrocriteria such as Technical aspects, Economic aspects, Reuse and Social aspects. After, each one of these macrocriteria can be detailed with specific subcriteria, so that:
- *intended use innovation* and *work duration* can be subcriteria of Technical aspects,
- *maintenance cost*, *net present value* and *payback period* can be subcriteria of Economic aspects,

- *impact on architectural value* and *physical impact* can be subcriteria of Reuse aspects,
- *human resources* can be a subcriterion of Social aspects.

The Multiple Criteria Hierarchy Process (MCHP-Corrente et al., 2012, 2013) is a methodology introduced in literature to take into account in an explicit way the hierarchy of criteria defining, in our context, an outranking relation $S_r$ for each node $g_r$ corresponding to some macrocriterion or subcriterion or also, at the root node, the overall evaluation, in the same hierarchy, so that $aS_rb$ means that $a$ is at least as good as $b$ on $g_r$. In this way, it is possible to get finer recommendations on the problem at hand by considering also specific aspects related to macro-criteria and sub-criteria at a time and not only overall evaluation related to conjoint consideration of all criteria simultaneously.

From a formal point of view, $g_o$ represents the root criterion; $g_r$ is a generic criterion in the hierarchy of criteria; $G$ is the set of all criteria in the hierarchy; $I_G$ is the set of the indices of all criteria in the hierarchy; $G_{EL} \subseteq G$ is the set composed of all elementary criteria, that is, the criteria at the bottom of the hierarchy (that therefore cannot be further detailed with lower level sub-criteria) and on which the alternatives are evaluated; $EL \subseteq I_G$ is the set of the indices of all elementary criteria; $g_t$, with $t \in EL$, denotes an elementary criterion, while, by the term non-elementary criterion, we refer to a criterion $g_r$ such that $r \in I_G \setminus EL$; finally, $E(g_r) \subseteq EL$ is the set of all elementary criteria descending from the non-elementary criterion $g_r$.

2.2 The hierarchical ELECTRE III method with interactions between criteria

Quite often, the criteria considered in decision problems are not mutually preferentially independent (Keeney and Raiffa, 1976; Wakker 1989), but they present a certain type of interactions. In the case of outranking methods and, in particular, for the ELECTRE ones we are applying in this paper, we distinguish between mutual-strengthening effect, mutual-weakening effect and antagonistic effects (Figueira et al., 2009). Elementary criteria $g_{t_1}$ and $g_{t_2}$ present a mutual-strengthening effect if the importance given to them together is greater than the sum of the importance given to them singularly; $g_{t_1}$ and $g_{t_2}$ present a mutual-weakening effect, if the importance given to them together is lower than the sum of their importance given to them singularly; finally, $g_{t_1}$ exercises an antagonistic effect over $g_{t_2}$, if the importance of $g_{t_1}$, being in favor of the outranking of a certain alternative $a$ over another alternative $b$, has to be lowered by the presence of $g_{t_2}$ being against the same outranking.

The ELECTRE III method belongs to the ELECTRE family (Figueira et al. 2013; Govindan and Jepsen, 2016). All ELECTRE methods are based on the construction of an outranking relation $S$ that is fulfilled by two alternatives $a$ and $b$, formally $aSb$, if a concordance and a non-discordance test are verified. The concordance test is verified if the majority of criteria is in favour of the outranking of $a$ over $b$, while the non-discordance test is verified if none of the remaining criteria opposes too strongly to such an outranking.

In the hierarchical ELECTRE III method with interactions, the concordance and discordance tests are dealt simultaneously by building, for each ordered pair of alternatives $(a, b) \in A \times A$, and for each non-elementary criterion $g_r$, a credibility index $\sigma_r(a, b)$ by means of the following steps:
1) For each elementary criterion $g_t \in E(g_r)$, a partial concordance index $\varphi_t(a, b)$ and a partial discordance index $d_t(a, b)$ are computed:

$$\varphi_t(a,b) = \begin{cases} 1 & \text{if} \quad g_t(b) - g_t(a) \leq q_t \quad (aS_t b) \\ \dfrac{p_t - [g_t(b) - g_t(a)]}{p_t - q_t} & \text{if} \quad q_t < g_t(b) - g_t(a) < p_t \quad (bQ_t a) \\ 0 & \text{if} \quad g_t(b) - g_t(a) \geq p_t \quad (bP_t a)^1 \end{cases}$$

$$d_t(a,b) = \begin{cases} 1 & \text{if} \quad g_t(b) - g_t(a) \geq v_t \\ \dfrac{[g_t(b) - g_t(a)] - p_t}{v_t - p_t} & \text{if} \quad p_t < g_t(b) - g_t(a) < v_t \\ 0 & \text{if} \quad g_t(b) - g_t(a) \leq p_t \end{cases}$$

where $q_t$, $p_t$ and $v_t$ are the indifference, preference and veto thresholds attached to $g_t$. In particular, $q_t$ represents the maximum difference between the performances of two alternatives on $g_t$, compatible with their indifference on the considered elementary criterion; $p_t$ represents the minimum difference between the performances of two alternatives on $g_t$, compatible with the preference of the better performing over the worse performing one; finally, $v_t$ is the minimum difference between the performances of two alternatives on $g_t$, incompatible with the outranking of one over the other. This means that, if $g_t(b) - g_t(a) \geq v_t$, then $a$ cannot outrank $b$ on any macro-criterion $g_r$ from which $g_t$ is descending, independently on their comparison on the other criteria (for more details see Roy et al., 2014).

Both $\varphi_t(a,b)$ and $d_t(a,b)$ belong to the interval [0,1] but they have a different interpretation: on one hand, $\varphi_t(a,b)$ measures how much $g_t$ is in favour of the outranking of $a$ over $b$. The higher $\varphi_t(a,b)$, the more $g_t$ is in favour of the considered outranking. On the other hand, $d_t(a,b)$ measures how much $g_t$ is against the outranking of $a$ over $b$. The higher $d_t(a,b)$, the more $g_t$ is against the considered outranking. More precisely, with respect to $\varphi_t(a,b)$,

- if $g_t(b) - g_t(a)$ is not greater than the indifference threshold $q_t$, then $g_t$ is fully in concordance with the outranking of $a$ over $b$, so that $\varphi_t(a,b)$ attains its maximum value, that is 1;
- if $g_t(b) - g_t(a)$ is not smaller than the preference threshold $p_t$, then $g_t$ is definitely not in concordance with the outranking of $a$ over $b$, so that $\varphi_t(a,b)$ attains its minimum value, that is 0;
- in all other cases, that is when $g_t(b) - g_t(a)$ is greater than the indifference threshold $q_t$ and smaller than the preference threshold $p_t$, then $g_t$ is partially in concordance with the outranking of $a$ over $b$, so that $\varphi_t(a,b)$ takes a value between 0 and 1, decreasing linearly from $\varphi_t(a,b) = 1$ (for $g_t(b) - g_t(a) = q_t$) to $\varphi_t(a,b) = 0$ (for $g_t(b) - g_t(a) = p_t$).

With respect to $d_t(a,b)$,

- if $g_t(b) - g_t(a)$ is not smaller than the veto threshold $v_t$, then $g_t$ is definitely in discordance with the outranking of $a$ over $b$, so that $d_t(a,b)$ attains its maximum value, that is 1;
- if $g_t(b) - g_t(a)$ is not greater than the preference threshold $p_t$, then $g_t$ is definitely not in discordance with the outranking of $a$ over $b$, so that $d_t(a,b)$ attains its minimum value, that is 0;

---

[1] $aS_t b$ means that $a$ is at least as good as $b$ on $g_t$; $bQ_t a$ means that $b$ is weakly preferred (that is , preferred with wone hesitation) to $a$ on $g_t$, while $bP_t a$ means that $b$ is strictly preferred to $a$ on $g_t$.

- in all other cases, that is when $g_t(b) - g_t(a)$ is greater than the preference threshold $p_t$, and smaller than the veto threshold $v_t$, then $g_t$ is partially in discordance with the outranking of $a$ over $b$, so that $d_t(a,b)$ takes a value between 0 and 1, increasing linearly from $d_t(a,b) = 0$ for $(g_t(b) - g_t(a) = p_t)$ to $d_t(a,b) = 1$ (for $g_t(b) - g_t(a) = v_t$).

2) After defining the importance $w_t$ of the elementary criteria $g_t$, the coefficients $w_{t_1 t_2}$ representing the mutual-weakening and mutual-strengthening effects between $g_{t_1}$ and $g_{t_2}$ and the coefficient $w'_{t_1 t_2}$ representing the antagonistic effect exercised by criterion $g_{t_2}$ over $g_{t_1}$, a partial concordance index $C_r(a,b)$ is computed for each non-elementary criterion $g_r$ and for each ordered-pair of alternatives $(a,b) \in A \times A$:

$$C_r(a,b) = \frac{1}{W_r(a,b)} \Bigg[ \sum_{t_1 \in \bar{C}(bPa) \cap E(g_r)} w_{t_1} \varphi_{t_1}(a,b)$$

$$+ \sum_{t_1, t_2 \in \bar{C}(bPa) \cap E(g_r)} w_{t_1 t_2} \min\left(\varphi_{t_1}(a,b), \varphi_{t_2}(a,b)\right)$$

$$- \sum_{\substack{t_1 \in \bar{C}(bPa) \cap E(g_r) \\ t_2 \in C(bPa) \cap E(g_r)}} w'_{t_1 t_2} \min\left(\varphi_{t_1}(a,b), \varphi_{t_2}(a,b)\right) \Bigg]$$

where $C(bHa)$ denotes the set of elementary criteria such that $bHa$, $H \in \{S, Q, P\}$, $\bar{C}(bHa)$ represents the complement of $C(bHa)$ and

$$W_r(a,b) = \Bigg[ \sum_{t_1 \in \bar{C}(bPa) \cap E(g_r)} w_{t_1} + \sum_{t_1, t_2 \in \bar{C}(bPa) \cap E(g_r)} w_{t_1 t_2} \min\left(\varphi_{t_1}(a,b), \varphi_{t_2}(a,b)\right) -$$

$$\sum_{\substack{t_1 \in \bar{C}(bPa) \cap E(g_r) \\ t_2 \in C(bPa) \cap E(g_r)}} w'_{t_1 t_2} \min\left(\varphi_{t_1}(a,b), \varphi_{t_2}(a,b)\right) \Bigg].$$

$C_r(a,b)$ is a value belonging to the interval [0,1] and it represents how much elementary criteria descending from $g_r$ are in favour of the outranking of $a$ over $b$. The higher $C_r(a,b)$, the more elementary criteria descending from $g_r$ are in favour of this outranking.

3) For each ordered pair of alternatives $(a,b) \in A \times A$ and for each non-elementary criterion $g_r$, the credibility index is therefore computed as follows:
$$\sigma_r(a,b) = C_r(a,b) \prod_{t \in E(g_r):\, d_t(a,b) > C_r(a,b)} \frac{1 - d_t(a,b)}{1 - C_r(a,b)}.$$
$\sigma_r(a,b)$ represents the credibility of the outranking of $a$ over $b$ and it is equal to the value $C_r(a,b)$ reduced in case some elementary criteria descending from $E(g_r)$ are opposing to this outranking. In particular, $\sigma_r(a,b) = C_r(a,b)$ iff none of the elementary criteria in $g_r$ opposes to the outranking of $a$ over $b$ and $\sigma_r(a,b) = 0$ if at least one of the elementary criteria in $E(g_r)$ strongly opposes to the considered outranking.

4) On the basis of the credibility indices $\sigma_r(a,b)$ computed in the previous step, for each non-elementary criterion $g_r$ two complete preorders (that is, a complete ranking admitting ex-aequo) of the alternatives are obtained by means of two specific procedures called ascending and descending distillation and, finally, a partial preorder (that is, a ranking admitting ex-aequo and also incomparability) of the same alternatives is computed as the intersection of the complete preorders found above.

In particular, with respect to a criterion $g_r$ and a pair of alternatives $a$ and $b$, one has
- a preference relation, denoted by $aP_rb$, if $aS_rb$ and not$(bS_ra)$ (that is, $a$ outranks $b$, but $b$ does not outrank $a$ with respect to $g_r$),
- an indifference relation, denoted by $aI_rb$, if $aS_rb$ and $bS_ra$ (that is, $a$ outranks $b$ and $b$ outranks $a$ with respect to $g_r$), and
- an incomparability relation, denoted by $aR_rb$, if not$(aS_rb)$ and not$(bS_ra)$ (that is, $a$ does not outrank $b$ and $b$ does not outrank $a$ with respect to $g_r$), (for more details see Corrente et al. 2017).

## 2.3 The hierarchical and imprecise Simos-Roy-Figueira (SRF) and the new SRF-II method

As explained in the previous section, the application of the hierarchical ELECTRE III method with interaction between criteria involves the knowledge of several parameters: one weight $w_t$ for each elementary criterion $g_t$, one coefficient $w_{t_1 t_2}$ for each pair of elementary criteria $g_{t_1}$ and $g_{t_2}$ for which there is a mutual-strengthening effect or mutual-weakening effect and one coeffient $w'_{t_1 t_2}$ for each ordered pair of criteria $g_{t_1}$ and $g_{t_2}$ such that one, $g_{t_2}$, exercises an antagonistic effect over the other $g_{t_1}$. However, asking the DM of providing directly all these technical parameters is meaningless since it involves a great cognitive effort from her part and the obtained result would be no reliable. To get the weights of the elementary criteria, a modification of the SRF method (Figueira and Roy, 2002) considering a hierarchy of criteria and an imprecise preference information from the part of the DM has been proposed in Corrente et al. (2017). For each non-elementary criterion $g_r$, the imprecise SRF method is applied to the set of criteria $G_r = \{g_{(r,1)}, \ldots g_{(r,n(r))}\}$ composed of the criteria descending from $g_r$ and being sited at the level immediately below it. The SRF elicitation procedure proceeds by asking the DM to:
1. Rank order the criteria in $G_r$ from the least important to the most important with the possibility of some ex-aequo between them;
2. Put some blank cards (an exact number $e_r$ or an interval of possible values $[e_r^l, e_r^u]$) between two successive subsets of criteria to increase the difference of importance between the criteria in these sets[2];
3. Define the ratio (an exact value $z$ or an interval of possible values $[z^l, z^u]$) between the weight of the most important criteria and the weight of the least important ones.

For the basic case in which only an exact number of cards is given between two successive subsets of criteria as well as an exact value z is given to the ratio between the weights of the most important and least important criteria, the non-normalized weight for criteria in the s-th level of q levels with increasing importance is given by (Corrente et al. 2016)

$$w_s = \frac{\sum_{r=1}^{s-1}(e_r+1)(z-1)}{\sum_{r=1}^{q-1}(e_r+1)} + 1.$$

Regarding the last question to the DM in the SRF elicitation procedure, in this paper, we propose an improvement to aid the DM to understand better the required information in order to get a more consistent and solid elicitation of the weights assigned to criteria.

---

[2] Let us observe that no blank cards between two successive subsets of criteria does not mean that the criteria in these sets have the same importance but that the difference of importance between them is minimal.

Indeed, while in step 2. the DM is asked to provide an information in terms of difference of importance between criteria placed in consecutive subsets, in step 3. the DM is asked to provide an information of different nature being the ratio (not the difference) between the importance of criteria sited in the extreme subsets. In other terms, the DM is asked to specify how many times the most important criteria are more important than the least important ones. Providing such a preference can be problematic for the DM and, for such a reason, in this paper we replaced point 3. above with the following one, introducing the new concept of SFR-II:

4. define the number of blank cards (an exact value or an interval of possible values) that should be included between a zero-level of importance and the least important criterion.

Let us discuss the nature and the reasons of asking the z value and the blank cards between the criterion 0 and the least important criterion. In Figueira and Roy (2002) a decision problem with four criteria ranked in increasing order of importance, let us say $g_1, g_2, g_3$ and $g_4$, is considered. Moreover, it is supposed that the DM considers the same difference of importance between each criterion and the following one. A first possible representation of this information in terms of cards is not inserting any blank card between the cards representing each criterion and giving to each criterion $g_r, r = 1,2,3,4$, a weight equal to the number of cards between the card corresponding to the least important criterion and the card corresponding to the same criterion $g_r$, so that weights $w_1 = 1, w_2 = 2, w_3 = 3$ and $w_4 = 4$ are given to $g_1, g_2, g_3$ and $g_4$, respectively. This is the procedure proposed by Simos (1990), which assigns a ration $z = \frac{w_4}{w_1} = 4$ between the weights of the most and the least important criterion. Observe, however, that the value of z is assigned implicitly, automatically and without taking into account the opinion of the DM that could prefer weights $w_1 = 3, w_2 = 4, w_3 = 5$ and $w_4 = 6$ or $w_1 = 6, w_2 = 7, w_3 = 8$ and $w_4 = 9$ that cannot be obtained with the Simos procedure. For this reason, Figueira and Roy (2002) proposes to ask the DM to supply the value of z. In this way, in the above example, with the same disposition of blank and non blank cards (in fact, in the considered example, only non-blank cards),

- if the DM gives the information $z = 4$, one obtains the weights $w_1 = 1, w_2 = 2, w_3 = 3$ and $w_4 = 4$,
- if the DM gives the information $z = 2$, one obtains the weights $w_1 = 3, w_2 = 4, w_3 = 5$ and $w_4 = 6$,
- if the DM gives the information $z = \frac{3}{2}$, one obtains the weights $w_1 = 6, w_2 = 7, w_3 = 8$ and $w_4 = 9$.

Observe that the same weights obtained with $z = 2$ could be assessed with the information that between the "zero level" and the least important criterion $g_1$ there are two blank cards. Analogously, five blank cards between the "zero level" and $g_1$ are equivalent to $z = \frac{3}{2}$ and gives the same weights. In general, denoting by $e_0$ the number of blank cards between the "zero level" and $g_1$, the following relation holds between z and b: $z = \frac{e_0+4}{e_0+1}$. Moreover, in general, in terms of $e_0$ rather than z, the non-normalized weight for criteria in the s-th level of q levels with increasing importance, is given by

$$w_s = \sum_{r=0}^{s-1}(e_r + 1).$$

Consequently, for the general case, there is the following relation between the values of z and $e_0$:

$$z = \frac{\sum_{r=1}^{q-1}(e_r + 1) + e_0 + 1}{e_0 + 1}.$$

From a behavioural point of view, we observed that the DM was more comfortable in applying the SRF-II method with $e_0$ rather than with $z$, that is, with step 4. instead of step 3. The reason is that supplying information in terms of $e_0$ rather than in terms of $z$, the DM is asked to provide in steps 2. and 4. a preference information of the same nature and that has, consequently, analogous interpretation. In this perspective we believe that our case study proves that the reformulation of the SRF method in terms of $e_0$ is a convenient and fruitful innovation for the whole methodology that can be applied advantageously in decision problems in any domain, not only in the ambit of architecture and urban and territorial planning.

As to the possible interactions between criteria as well as regarding the antagonistic effect between them, the DM is only asked to provide the type of such an interaction, that is, she has only to say if she retains that there is a mutual-weakening or a mutual-strengthening effect between two elementary criteria or that there exists an antagonistic effect exercised by an elementary criterion over another one.

## 2.4 Stochastic Multicriteria Acceptability Analysis

In general, more than one vector of parameters (weights of criteria and coefficients representing mutual-weakening, mutual-strengthening and antagonistic effects) could be compatible with the preference information provided by the DM. For this reason, providing a final recommendation on the problem at hand by using only one of these sets is meaningless and arbitrary to some extent. The Stochastic Multicriteria Acceptability Analysis (SMAA) avoids such a choice considering all vectors of compatible parameters and, therefore, providing robust recommendations on the considered problem. In particular, the application of the SMAA methodology to the hierarchical ELECTRE III method gives information in statistical terms (for more details see Corrente et al. (2017)) providing:

- for each non-elementary criterion $g_r$ and for each ordered pair of alternatives $(a, b) \in A \times A$, the probability of preferences $(P_r(a, b))$, indifference $(I_r(a, b))$ and incomparability $(R_r(a, b))$ between $a$ and $b$ on $g_r$;
- for each non-elementary criterion $g_r$ and for each alternative $a \in A$, the mean number of alternatives $b$ such that $a$ is at least as good as $b$ on $g_r$;
- for each non-elementary criterion $g_r$ and for each alternative $a \in A$, the mean number of alternatives $b$ such that $b$ is at least as good as $a$ on $g_r$.

## 2.5 The concept of adaptive reuse

In a context of increasing amount of abandoned buildings, the adaptive reuse represents a valuable practice to exploit this potential asset. In the 20[th] century this phenomenon begins to be identified as a creative discipline for preserving the cultural heritage and for tackling the huge social, technological and environmental changes (Plevoets and Van Cleempoel, 2013), assuming that buildings, areas, districts and sites are not static entities designed simply for one use during their life cycle.

The theory of the adaptive reuse emerged first as a practice, then as a theory, of introducing a new content in an existing container (i.e. building, infrastructure, area), paying particular attention to the needs of the society and following the principle of the maximum conservation and the minimum transformation (Robiglio, 2016). Furthermore, this practice should not involve heavy work and changes to the existing buildings, but it exploits the heritage as opportunities to ameliorate disused buildings into new items with new purposes (Bullen and Love, 2011a). Buildings become obsolete but they continue to represent a value or a symbol for the place or for the community, they keep an intrinsic memory, which is difficult to wipe up without opposition (Fiorani et al., 2017).

Address the problem of reuse in a perspective of adaptive reuse (Günçea and Mısırlısoy, 2015, Young and Chan, 2012) represents an increasing strategy for existing buildings: the extension of their life gives benefits to the investors from an economic point of view (Dyson et al 2016, Douglas 2006) and contributes to global climate protection and emission reduction (Conejos et al, 2014, Elefante 2007). More in detail, a considerable amount of literature states that the adaptation of a building or an area is cheaper than creating a new one (Douglas 2006, Bullen and Love, 2011b, Remøy and Van der Voordt, 2007, Kohler and Yang, 2007). Particularly interesting, in a sustainable perspective, is that reuse is encouraged respect to demolition and reconstruction as it reduces the consumption of raw materials and energy used in the process, reduces waste and preserves portions of the urban landscape.

3. The case study

The Stock Exchange is located in the city centre of Turin, standing in a very accessible point (Figure 1).

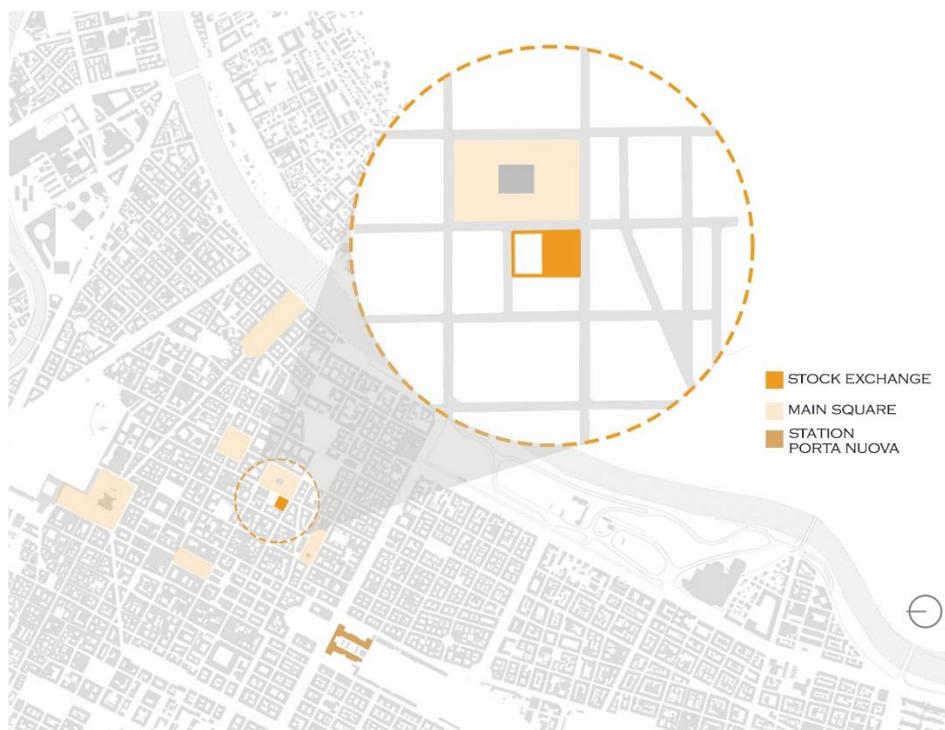

*Figure 1."Stock Exchange" building location in the city centre of Turin*

The building has been realized between 1952-57, as the new venue of the stock exchange of Turin according to the project of two well-known Italian architects, Aimaro Oreglia d'Isola and Roberto Gabetti. Moreover, the Stock Exchange is considered an iconic building since it represents the rebellion against the simplified forms free of ornaments, typical of the Modern Movement, in favour of the freedom of the styles, proper of the Italian taste of the 60s (Papuzzi, 2011). Two elements can be identified as its primary distinctive features: the great "Shouts Hall" and the peculiar pavilion vault covering the hall, which shows a size of about 40 m on side and it has been realized through a particular technical-constructive.

Concerning the "Shouts Hall", it can be described as the main centre of all the stock exchange activities and it develops on a surface of about 1500 sqm on a height of 17 m. Figure 2 shows on the left the construction of the vault and on the right side, the image of the "Shouts Hall" during trading.

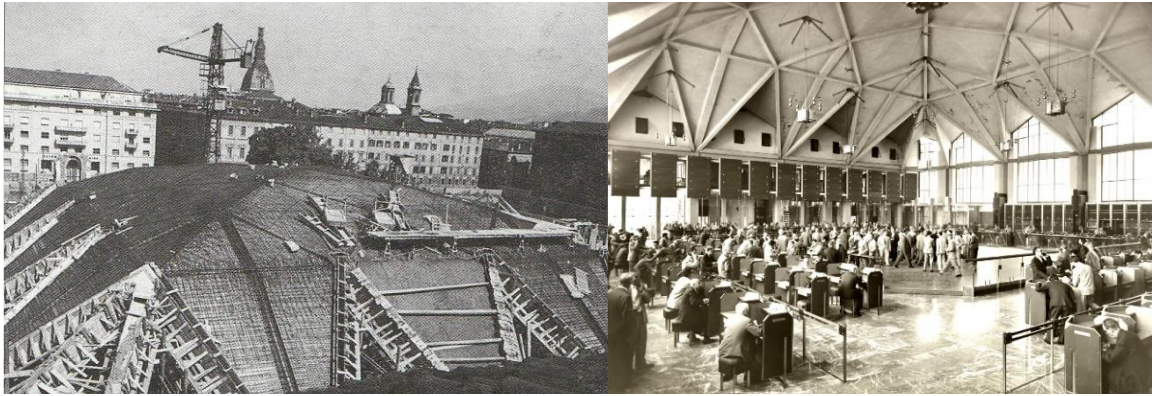

*Figure 2. Construction phases of the vault, 1955. (left, Alberto Papuzzi, 2011) and the «Shouts Hall» during trading (right, Chamber of Commerce of Turin documentation, www.to.camcom.it/ex-borsa-valori).*

The building also includes an office block, which develops on three levels and in the past hosted a wardrobe, a living room, some offices, meeting rooms and the apartment of the caretaker.

When the telematic exchanges replaced the shouted market in 1992, the "Stock Exchange" lost its function and it was abandoned in 2008.

Despite an attempted of project transformation in 2010 and a series of temporary artistic events since 2015, the building is currently without a function.

It is worth mentioning, that since the building has obtained the Title of "important artistic merit" issued by the Superintendence of Archeology, Fine Arts and Landscape for the Metropolitan City of Turin and since one of its two original authors, Aimaro Oreglia d'Isola, is currently alive, it is subject to the Copyright Law. This means that the building should be protected from deformation and mutilation and his author holds the right of paternity.

### 3.1 Definition of the project alternatives

In order to identify the most interesting transformation for the building, six alternative projects have been considered (Figure 3). Five of these are the hypotheses of transformation presented by students of the Master's degree program in Architecture Construction and City at the Politecnico di Torino, while the sixth alternative represents the project proposal commissioned to the Politecnico di Torino (Department of Architecture and Design) by the Piedmont Region and by the Chamber of Commerce of Turin, owner of the building.

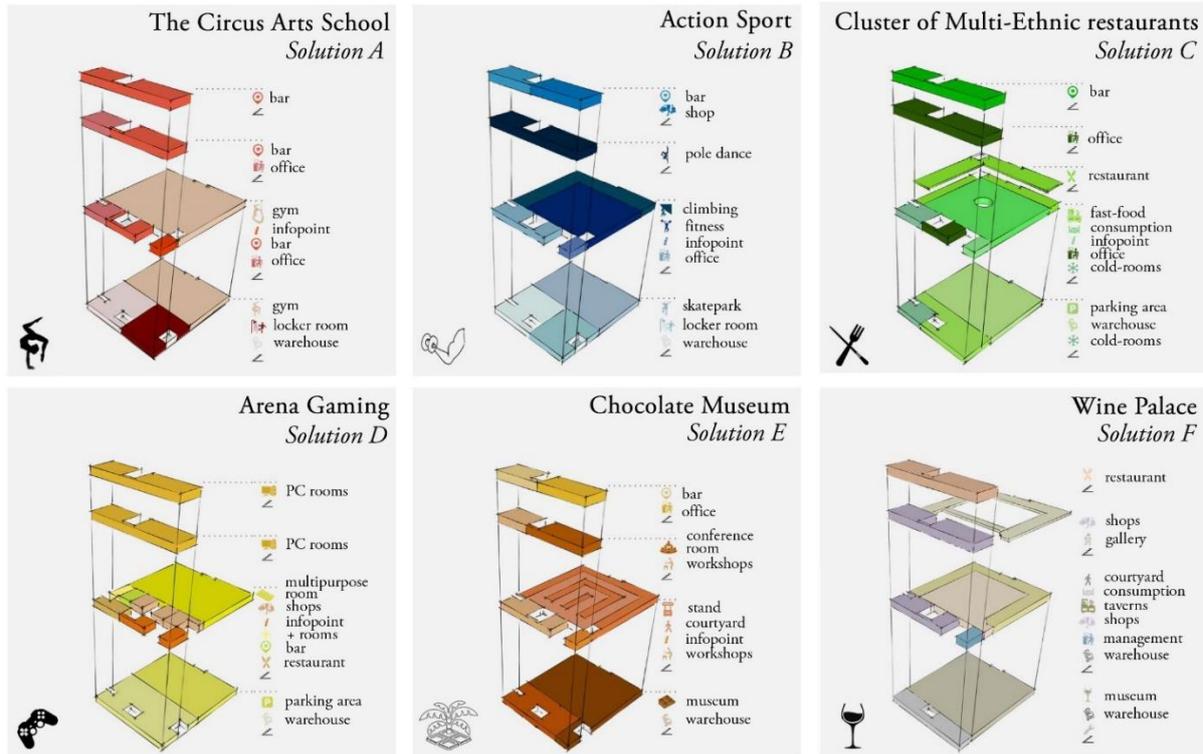

*Figure 3. Functional axonometric exploded views of the six designs alternatives*

From Figure 3 it is possible to see that the six alternative projects are heterogeneous. In particular, it is possible to distinguish two macro families:

1) *traditional functions* in line with the gastronomic and museal tradition of the city of Turin, such as the solutions C, E and F. In details:
    - solution C provides to transform the ancient stock market into a cluster of ethnic restaurants, conceived as fast food, following the growing presence of ethnic cultures and the consequent influence on the eating habits of Italians. The great high of the "Shouts Hall" is exploited with the addition of a mezzanine, on which an Italian restaurant could be placed. Moreover, in addition to the offices and the warehouse, a bar area is planned on the top floor of office block, which will also have access to the terrace that opens onto the Turin panorama;
    - solution E proposes to transform the "Stock Exchange" into a chocolate museum named Chocolate Island and located in the basement. The space of the "Shouts Hall" would be used for the layout of 24 commercial stands, whose arrangement will involve a partial, semi-reversible division of the single space. Spaces for didactical workshops, conference room, offices and a bar area would be inserted in the former office block;
    - solution F proposes to transform the building into a wine museum combined with taverns and a gourmet restaurant. The "Shouts Hall" would turn into a covered square used for visual narratives, events and around it the taverns are displaced. As the same as for the solution E, the museum and the spaces for didactical workshop would be located in the basement, while three shops and the gourmet restaurant would be hosted in the former office block.
2) *innovative solutions* proposing amusement activities, such as the solutions A, B and D. In particular:

- solution A proposes the new Circus Arts School. The large single free space of the "Shouts Hall" with its great height of 17 m looks perfect for the use of the flying trapeze, which needs to be fixed with a rope of at least 3 m, at 7.2 m from the ground, and for the insertion of a removable grandstand to welcome the public during the events. The basement level would be used as a gym for the various courses and the office block would house a bar area on three levels and an office area on the first and second level;
- solution B hypothesizes to insert functions mainly addressed to a young public, or a multi-sport center, equipped for extreme sports such as skateboarding, climbing, pole dance and a part of fitness gym. The potential of the height offered by the "Shouts Hall" is exploited with the inclusion of the climbing wall needing for a double height. The large space offered by the first floor and the basement is instead exploited by the placement of a fitness area and a skate park. Moreover, in the office block offices, would be placed a space dedicated to the pole dance, a bar area and a shop for the sale of the equipment necessary for the practice of the aforementioned sports.
- solution D is the Arena Gaming and follows the strongly growing international market of the e-games: electronic sports practiced at a competitive and professional level through video games. The "Stock Exchange" will house the new video games centre as a meeting place and dialogue on eSport. The former "Shouts Hall" would turn into the new multi-purpose hall in order to house the national and international tournaments or extra events. Furthermore, the office block would host three shops related to the e-games sector, while a bar, a restaurant area and PC rooms would be placed in the last two floors.

Concerning the interventions on the building, all projects involve: the restoration of the facades, the insertion of new plants system and of an acoustic and thermal insulation system to allow the use of the building and, finally, the insertion of the equipment necessary for the various activities.

4. Structuring the decision-aiding process

We proceeded to structure the subsequent decision-aiding process aimed at identifying the preferred solution according to the preferences of the DMs. Therefore, this section describes the decision-making process according to the methodological framework explained in section 2.

4.1 Criteria, sub-criteria and evaluation matrix

The alternative projects have been evaluated on a coherent, exhaustive and non-redundant set of criteria (Roy and Bouyssou, 1993). These criteria were defined on the basis of the features of the building and according to the literature concerning the synthesis of problems into criteria or clusters of criteria (Martin e Lagret, 2005, Abastante e Lami, 2013, Bottero et al.,2015, Abastante, 2016, Camoletto et al.,2017, Abastante et al., 2018)

A set of four macro-criteria has been defined, which are in turn composed of eight quali-quantitative sub-criteria as shown in Table 1. It is worth mentioning that these have been defined in the perspective of the property owner, the Chamber of Commerce of Turin.

*Table 1. Criteria, sub-criteria and related units of measure.*

| $G_N$ | Macro-Criteria | $g_n$ | Sub-criteria | Units of measure | Preference direction of $g_n$ |
|---|---|---|---|---|---|

| $G_T$ | Technical | $g_{T1}$ | Intended use innovation | Ordinal scale | ↑ |
|---|---|---|---|---|---|
| | | $g_{T2}$ | Work Duration | Months | ↓ |
| $G_E$ | Economic | $g_{E1}$ | Maintenance Cost | % | ↓ |
| | | $g_{E2}$ | Net Present Value | € | ↑ |
| | | $g_{E3}$ | Pay Back Period | Years | ↓ |
| $G_R$ | Reuse | $g_{R1}$ | Impact on architectural value | % | ↓ |
| | | $g_{R2}$ | Physical impact | Dichotomous scale | ↓ |
| $G_S$ | Social | $g_{S1}$ | Human resources | Numbers | ↑ |

The criteria have been organized into a hierarchical structure according to the MCHP (2.2). As we can see in Figure 4 the main issue, i.e. the choice of the transformation project, has been unpacked into four macro-criteria followed by the eight sub-criteria.

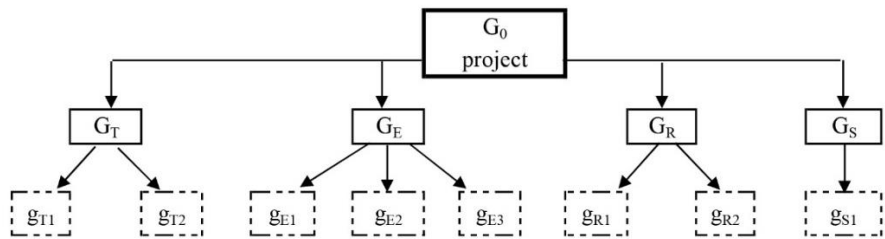

*Figure 4. Hierarchical structure of criteria*

A graphical representation with a brief description of the considered sub-criteria is set out in Table 2.

*Table 2. Graphical representation of the evaluation matrix of all design solution*

| Evaluation | Description |
|---|---|
| $G_{T1}$ [Ordinal scale]<br>5 MAXIMUM INNOVATION: $x_1 \leq 1$; $x_2 = 0$; $x_3 = 0$.<br>4 GOOD INNOVATION: $2 < x_1 \leq 4$; $x_2 = 0$; $x_3 = 0$.<br>3 AVARAGE INNOVATION: $0 < x_1 \leq 5$; $1 \leq x_2 \leq 3$; $x_3 = 0$.<br>2 LOW INNOVATION: $5 < x_1 \leq 10$; $3 < x_2 \leq 6$; $0 < x_3 \leq 2$.<br>1 NO INNOVATION: $x_1 > 10$; $x_2 > 6$; $x_3 > 2$. | The *intended use innovation* measures the degree of innovation, according to its diffusion at national ($x_1$), regional ($x_2$) and local ($x_3$) level. The values are expressed with an ordinal scale from 1 to 5, in which in comparing two options, the one with the highest performance is preferred to the other. |
| $G_{T2}$ [Months] (chart showing values 24 and 12) | The *work duration* assesses the time required for completion of the work. Since a limited duration of works is always desirable, the criterion must be minimized, or in the comparing two projects, the project with a shorter duration of works will be preferred to the other. |
| $G_{E1}$ [%] (chart showing values 5.2%, 5.1%, 5.0%, 4.6%, 4.1%) | The *level of maintenance* is defined as the incidence of the extraordinary maintenance costs charged to the owner on his/her total rent revenue, discounted to the tenth year. Also in this case the criterion is minimized. |
| $G_{E2}$ [Euro] (chart showing values 2.826.080, 2.640.840, 2.634.310, 2.380.320, 2.239.710, 1.827.780) | The *Net Present Value (NPV)* represents the sum of the cash flows (revenues-costs), generated by the investment and discounted at the initial time, in order to assess the profitability and feasibility of the intervention. Since, by definition, a project is feasible if the NPV is positive, the greater is its value, the greater are the chances of success, therefore the criterion is maximized. |
| **G**$_{E3}$<br><br>[Years]<br><br>11<br>10<br>9 | The *Payback Period (PBP)* is the time required for incoming (discounted) cash flows to equal outgoing cash flows.<br>Since it is a logic goal to recover the cost of the investment as soon as possible, the criterion should be minimized. |
| **G**$_{R1}$<br><br>9%<br>4%<br>3%<br>0%<br>[%] | The "*Shouts Hall*" is considered the distinctive value of the building, so that the criterion measures if the new projects decrease the free space of the "Shouts Hall" by splitting or inserting differentiated volumes within it, as a percentage of built volume on the total volume of the room. According to the concept of minimum transformation of the adaptive reuse, in the comparison in pairs the action with a lower impact will be preferred. |
| **G**$_{R2}$<br><br>Action Sport provides a climbing wall, which will cover part of a vault and consequently will be fixed to it.<br>YES<br><br>The Circus school plans to fix the trapezoid on the vault using a system of reticular beams.<br>NO<br><br>[Dichotomous scale] | The *physical impact* of the new project on the existing building is considered as a possible devaluation of the peculiar pavilion vault. The new project should not entail any impact on it with the inclusion of elements, which would damage its integrity and limit the possibility of admiring it in its entirety. Projects not impacting the building will be preferred. |

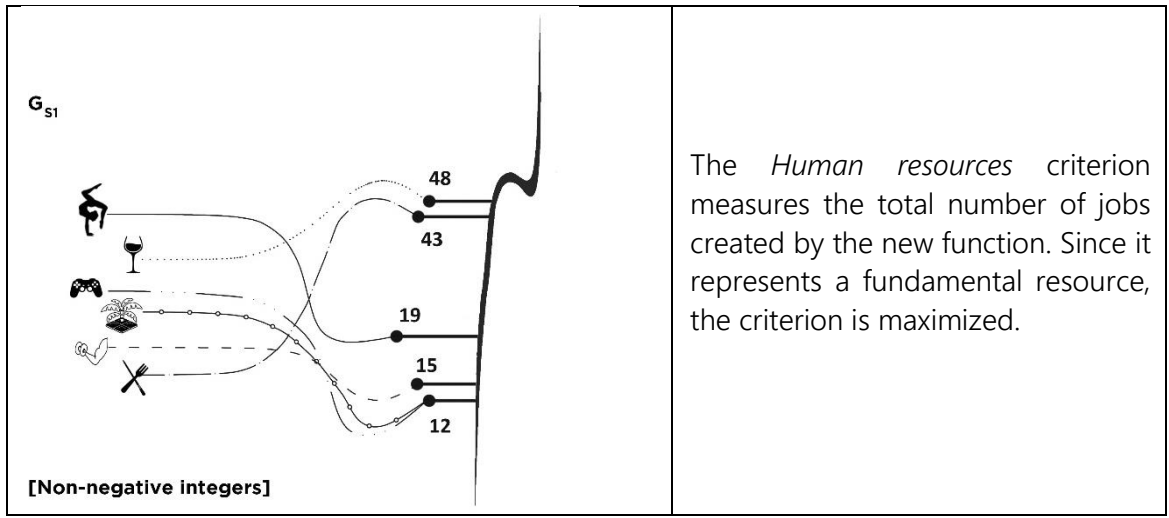

The *Human resources* criterion measures the total number of jobs created by the new function. Since it represents a fundamental resource, the criterion is maximized.

Table 3 summarises the performance matrix of the six projects respect to each elementary criterion.

*Table 3. Performance matrix*

|       | A         | B         | C         | D         | E         | F         |
|-------|-----------|-----------|-----------|-----------|-----------|-----------|
| $g_{T1}$ | 2         | 4         | 5         | 5         | 2         | 5         |
| $g_{T2}$ | 12        | 24        | 24        | 24        | 12        | 24        |
| $g_{E1}$ | 4,1%      | 4,6%      | 5,1%      | 5,0%      | 5,1%      | 5,2%      |
| $g_{E2}$ | 1.827.779 € | 2.239.710 € | 2.640.840 € | 2.634.312 € | 2.380.323 € | 2.826.078 € |
| $g_{E3}$ | 10        | 11        | 10        | 10        | 10        | 9         |
| $g_{R1}$ | 0%        | 0%        | 4%        | 3%        | 9%        | 9%        |
| $g_{R2}$ | yes (1)   | yes (1)   | no (0)    | no (0)    | no (0)    | no (0)    |
| $g_{S1}$ | 19        | 15        | 43        | 12        | 12        | 48        |

## 4.2 Decision makers and discrimination thresholds

For the application of the methodology, the DMs are asked to provide different types of preference information. The DMs are here two members of the project team that acts in the real process. Four meetings were needed over the course of about two months to define the complete picture of the preferential information.

First, the project alternatives and the evaluation criteria were exhaustively presented to them. Second, they were asked to express the discrimination thresholds; third they were asked to express information about possible mutual-strengthening and mutual-weakening effects, as well as about the possible antagonistic effects between some elementary criteria. Fourth, they provide some information by the imprecise SRF-II method on each subset of criteria and each sub-criterion.

Regarding the discrimination thresholds, Table 4 shows for each criterion the indifference $q$, preference $p$ and the veto $v$ threshold expressed by the DMs.

*Table 4. Discrimination thresholds expressed by DMs*

| $g_n$ | Units of measure | Scale | *indifference* | *preference* | *veto* |
|-------|------------------|-------|----------------|--------------|--------|

| | | | | | | |
|---|---|---|---|---|---|---|
| $g_{T1}$ | Ordinal scale | (1-5) | | 1 | 2 | - |
| $g_{T2}$ | Months | (1-∞) | | 12 | 24 | 36 |
| $g_{E1}$ | % | (1-100%) | | 0.1% | 0.3% | 17% |
| $g_{E2}$ | € | (0-∞) | | 100.000 | 200.000 | 1.000.000 |
| $g_{E3}$ | Years | (1-∞) | Until 5 | 1 | 4 | 15 |
| | | | Over 5 | | 2 | |
| $g_{R1}$ | % | (1-100%) | | 0.5% | 1% | 10% |
| $g_{R2}$ | Dichotomous scale | Yes/No | | - | - | - |
| $g_{S1}$ | Numbers | (1-∞) | Until 20 | 1 | 2 | 10 |
| | | | Over 20 | 4 | 7 | |

It is worth mentioning that the methodology allows the possibility to express thresholds depending on the performances of the alternatives on the criterion at hand. This was relevant for the definition of some thresholds, such as those on criterion $g_{E3}$, in which the DMs according to the payback period expressed a preference threshold depending on the evaluations:
- if $G_{E3}(a) \leq 5$, then $p_{E3}(a) = 4$ (to be strictly preferred to $a$ an alternative $b$ must have a payback period shorter than $a$ of at least 4 years)
- if $G_{E3}(a) > 5$, then $P_{E3}(a) = 2$ (to be preferred to $a$ an alternative $b$ must have an a payback period shorter than $a$ of at least 2 years).

Moreover, the sub-criterion $g_{R2}$, being measured on a dichotomous scale, does not require any preference and indifference thresholds.

### 4.3 Interaction between the considered criteria

A discussion is given concerning the possibility that two criteria can interact each other presenting a mutual-strengthening effect, a mutual-weakening effect or an antagonistic effect. To help DMs to correctly understand and explain these reciprocities, some possible interactions between the set of criteria have been presented them, on which, based on their technical knowledge, they were free to state an agreement or disagreement and present further interactions. As shown in Table 5, they expressed 4 strengthening effects and one weakening and antagonistic effects, which numerical value was later determined mathematically.

*Table 5. Description of the interaction between the criteria*

| $g_n$ | Criterion 1 | $g_n$ | Criterion 2 | Interaction | Description |
|---|---|---|---|---|---|
| $g_{T1}$ | Intended use innovation | $g_{T2}$ | Work Duration | Strengthening effect | If a project is characterized both by a high innovation of intended use and by a low work duration, the importance of the two criteria together must be considered greater than the sum of their importance when they are considered alone |
| $g_{R1}$ | Impact on architectural value | $g_{R2}$ | Physical impact | Strengthening effect | If a project is characterized by a low impact on the architectural value of the building and a low physical impact, the importance of the two |
| | | | | | criteria must be considered greater than the sum of their importance when they are considered alone |
| $g_{E2}$ | NPV | $g_{R1}$ | Physical impact | Antagonism effect | If a project has good performance on the income compared to another, but provides a high impact on the building, the contribution of the Net Present Value criterion must be considered less than its weight. |
| $g_{E2}$ | NPV | $g_{S1}$ | Human resources | Strengthening effect | If a project has good performance on the income compared to another and also provides more jobs, the importance of the two criteria must be considered greater than the sum of their importance when they are considered alone. |
| $g_{T1}$ | Intended use innovation | $g_{E3}$ | PBP | Strengthening effect | If a project is characterized both by a high innovation of intended use and by a short time of return of the investment with respect to the other, the importance of the two criteria must be considered greater than the sum of their importance when they are considered alone. |
| $g_{E3}$ | PBP | $g_{E2}$ | NPV | Weakening effect | If a project is characterized by a short initial investment return time, it will be easy for it to also present a positive NPV, so the importance of the two criteria must be considered lower than the sum of their importance when they are considered alone. |

### 4.4 Prioritization of all criteria/sub-criteria and the new SRF-II method

The last essential preferential information required by the methodology, entails the application of the SRF-II method in order to define the set of feasible weights of the criteria and sub-criteria.
In order to simplify required information reducing the cognitive burden for the DMs from whom we had to elicit the weights for considered criteria, the SRF-II method based on the number of blank cards $e_0$ between the "zero - level" and the least important criteria introduced in Section 2.3 was applied.
Since every DM has its own opinions and preferences, it might be interesting to observe how the results of the application could be different by changing the input values. In light of this, the SRF-II method was first applied separately for each DM and then they were required to interact with each other in order to provide a common classification of the criteria and of the sub-criteria. Here we provide the set of the common preferences, namely:

- With respect to the first hierarchical level criteria ($G_T, G_E, G_R$ and $G_S$), the experts specify that since the building in question is a public good, the social and economic factors represent the priority of the transformations. Accordingly, they expressed that $G_T$ is less important than $G_R$, which in turn is less important than $G_S$ that is overtaken by $G_E$. No blank cards had been inserted between $G_T$ and $G_R$, while the number of the blank cards between $G_R$ and $G_S$ belongs to the interval [2-3] and the ones between $G_S$ and $G_E$ belongs to the interval [0-1]. Moreover, they decided to insert 3 blank cards between the less important criteria ($G_T$) and the "zero - level" ($e_0 = 3$) and, therefore, the ratio z is in the interval [9-11].
- Considering the macro-criterion $G_T$, they express that $g_{T1}$ is not a binding factor but an added value to the project, therefore ranked after the criterion $g_{T2}$. The blank cards inserted between them belongs to the interval [2-4]. They inserted 1 blank card between the criterion $g_{T1}$ and the "zero level" ($e_0 = 1$) and, therefore, the ratio z belongs to the interval [5-7].
- With respect to the macro-criterion $G_E$, they ranked the criteria $g_{E2}$ and $g_{E3}$ in *ex-aequo* at the first level, since the NPV and the PBP are factors that highlight the solidity of the project. The criterion $g_{E1}$ is less troubling if the project scores a positive NPV, so it gets the last rank. The blank cards inserted between $g_{E1}$ and {$g_{E2}, g_{E3}$} are 2. They decided that at least 1 and at most 3 blank cards should be added between the less important criteria ($g_{E1}$) and the "zero - level" ($e_0 \in [1,3]$) and, therefore, the ratio z belongs to the interval [5-7].
- Considering the macro-criterion $G_R$, DMs ranked $g_{R1}$ at the last place after $g_{R2}$. The latter represents the most relevant criterion since the vault represents the distinctive element of the building and therefore should not be devalued. The blank cards inserted between them belongs to the interval [0-1]. They inserted 1 blank card between the criterion $g_{R1}$ and the "zero – level" ($e_0 = 1$); consequently, the ratio z belongs to the interval [3-4].
- With respect to the macro-criterion $G_S$, since it provides only one sub-criterion there is no need to express a preference ranking.

5. Final results

In light of the preference information provided by the DMs and presented in the previous section, we checked if there exists at least one vector of parameters compatible with this preference and, indeed, it was the case. For the purpose of robust recommendations, the SMAA methodology had been applied to summarize the results of the hierarchical ELECTRE III method with interactions, obtained sampling 10,000 compatible vectors. Consequentially, the methodology provides several different partial pre-orders of alternatives compatible with the preferences expressed by the DM, with the relative frequencies with which they occurred. For each considered partial pre-order, we computed the barycenter of the vectors of parameters restoring it; this barycenter represents the average preferences provided by the DM for which the considered pre-order is obtained. Moreover, together with the most frequent partial pre-orders, for each non-elementary criterion and for the root criterion the frequency with which an alternative is preferred, indifferent or incompatible to another has been provided. Lastly, the application of the applied methodology permits to get for each alternative *a*, the mean number of alternative outranked by *a* and the mean number of alternatives outranking *a*, not only at comprehensive level but also considering one of the four macro-criteria singularly.

We report here the results at the comprehensive level obtained considering the preference provided by the two DMs together; similar results concerning the other four macro-criteria highlighted in the hierarchy are provided in the appendix.

Computing one partial pre-order for each of the 10,000 sampled compatible vectors, 13 different partial pre-orders are obtained and in Figures 5-7 we show the three obtained more frequently. In particular, these three pre-orders appear with frequencies of the 67.34%, 12.54% and 10.67% of the cases, respectively. Consequently, the other ten partial pre-orders are obtained, together, in the remaining 9.45% of the considered cases.

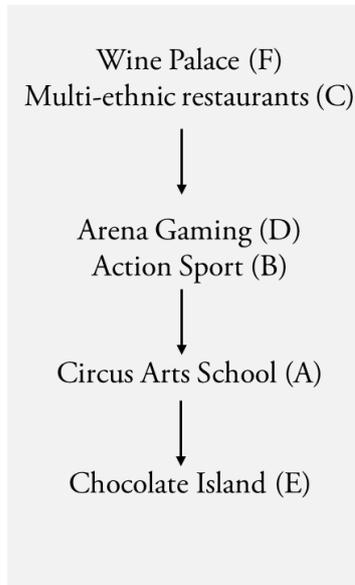 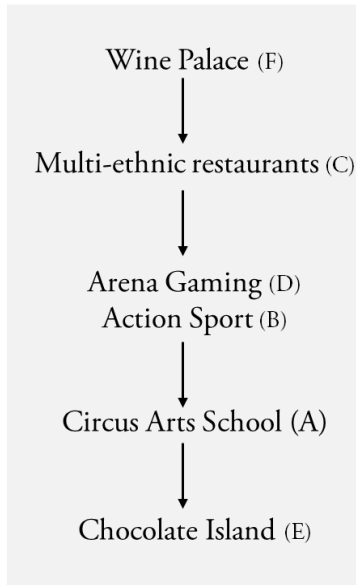 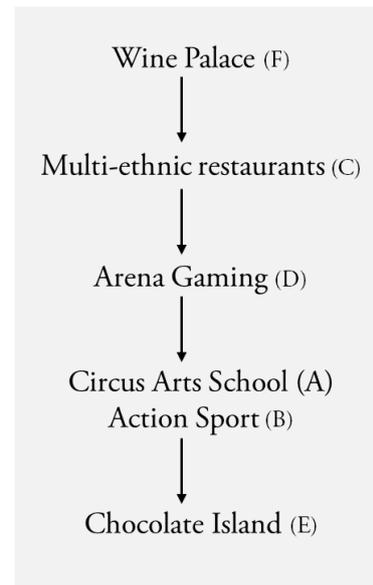

*Figure 5. Most frequent (67,34%) partial pre-order at comprehensive level*

*Figure 6. Second most frequent partial pre-order obtained with a frequency of 12,54%*

*Figure 7. Third most frequent (10,67 %) partial pre-order at comprehensive level*

Looking at the mentioned figures, one can observe that Wine Palace is always in the first place in the three rankings and it has to share the first position with Multi-ethnic restaurants in the most frequent partial pre-order. Analogously, Chocolate Island is the less preferred option in the three most frequent rankings.

Looking at Table 6, one can observe that Wine Palace project is preferred to all other alternative projects, apart from Multi-Ethnic Restaurant project, in all cases. Looking only at these two projects, quite often they are indifferent (70.85% in Table 5), Wine Palace is preferred to Multi-Ethnic Restaurant with a frequency of the 25.28%, while the opposite is true in the remaining cases (3.87% in Table 6).

*Table 5. Frequency of indifference*

|   | A | B | C | D | E | F |
|---|---|---|---|---|---|---|
| A | 0 | 17,44 | 0 | 0 | 0 | 0 |
| B | 17,44 | 0 | 0 | 80,07 | 0 | 0 |
| C | 0 | 0 | 0 | 0 | 0 | 70,85 |
| D | 0 | 80,07 | 0 | 0 | 0 | 0 |
| E | 0 | 0 | 0 | 0 | 0 | 0 |
| F | 0 | 0 | 70,85 | 0 | 0 | 0 |

Table 6. Frequency of preference

|   | A | B | C | D | E | F |
|---|---|---|---|---|---|---|
| A | 0 | 0 | 0 | 0 | 100 | 0 |
| B | 82,56 | 0 | 0 | 0,49 | 100 | 0 |
| C | 100 | 100 | 0 | 100 | 100 | 3,87 |
| D | 97,36 | 17,29 | 0 | 0 | 100 | 0 |
| E | 0 | 0 | 0 | 0 | 0 | 0 |
| F | 100 | 100 | 25,28 | 100 | 100 | 0 |

Table 7. Frequency of incomparability

|   | A | B | C | D | E | F |
|---|---|---|---|---|---|---|
| A | 0 | 0 | 0 | 2,64 | 0 | 0 |
| B | 0 | 0 | 0 | 2,15 | 0 | 0 |
| C | 0 | 0 | 0 | 0 | 0 | 0 |
| D | 2,64 | 2,15 | 0 | 0 | 0 | 0 |
| E | 0 | 0 | 0 | 0 | 0 | 0 |
| F | 0 | 0 | 0 | 0 | 0 | 0 |

As already underlined above, we computed, for each obtained partial pre-order, the barycenter of the vectors of parameters for which the application of the ELECTRE III method provided the considered ranking. Looking at the values in Table 8, one can observe that the weight given to the elementary criteria is almost the same in the three barycenters. What is making the difference so that the partial pre-orders are different, is the value assigned to the coefficients representing the interaction and antagonistic effects. Indeed, in the three barycenters, the interaction effect presenting the highest value is that one between Net Present Value $(g_{E_2})$ and Human Resources $(g_{S_1})$.

Table 8. Barycenters of the vectors of parameters giving a certain partial pre-order at comprehensive level

| Most frequent ranking | | | | | | | |
|---|---|---|---|---|---|---|---|
| $G_T$ | | $G_E$ | | | $G_R$ | | $G_S$ |
| 0,068 | | 0,445 | | | 0,132 | | 0,355 |
| $g_{T1}$ | $g_{T2}$ | $g_{E1}$ | $g_{E2}$ | $g_{E3}$ | $g_{R1}$ | $g_{R2}$ | $g_{S1}$ |
| $W_{11}$ | $W_{12}$ | $W_{21}$ | $W_{22}$ | $W_{23}$ | $W_{31}$ | $W_{32}$ | $W_{41}$ |
| 0,108 | 0,038 | 0,070 | 0,147 | 0,147 | 0,067 | 0,115 | 0,308 |

| Strengthening | | | | Weakening | Antagonism |
|---|---|---|---|---|---|
| $g_{T1}, g_{T2}$ | $g_{T1}, g_{E3}$ | $g_{E2}, g_{S1}$ | $g_{R1}, g_{R2}$ | $g_{E2}, g_{E3}$ | $g_{E2}, g_{R3}$ |
| $W_{11\_12}$ | $W_{11\_23}$ | $W_{22\_41}$ | $W_{31\_32}$ | $W_{22\_23}$ | $W_{22\_32}$ |
| 0,131 | 0,194 | 0,353 | 0,090 | -0,032 | 0,040 |

| Second most frequent ranking | | | | | | | |
|---|---|---|---|---|---|---|---|
| $G_T$ | | $G_E$ | | | $G_R$ | | $G_S$ |
| 0,068 | | 0,445 | | | 0,132 | | 0,355 |
| $g_{T1}$ | $g_{T2}$ | $g_{E1}$ | $g_{E2}$ | $g_{E3}$ | $g_{R1}$ | $g_{R2}$ | $g_{S1}$ |
| $W_{11}$ | $W_{12}$ | $W_{21}$ | $W_{22}$ | $W_{23}$ | $W_{31}$ | $W_{32}$ | $W_{41}$ |
| 0,107 | 0,037 | 0,070 | 0,149 | 0,149 | 0,065 | 0,114 | 0,309 |

| Strengthening | | | | Weakening | Antagonism |
|---|---|---|---|---|---|
| $g_{T1}, g_{T2}$ | $g_{T1}, g_{E3}$ | $g_{E2}, g_{S1}$ | $g_{R1}, g_{R2}$ | $g_{E2}, g_{E3}$ | $g_{E2}, g_{R3}$ |
| $W_{11\_12}$ | $W_{11\_23}$ | $W_{22\_41}$ | $W_{31\_32}$ | $W_{22\_23}$ | $W_{22\_32}$ |
| 0,099 | 0,077 | 0,179 | 0,090 | -0,046 | 0,072 |

| Third most frequent ranking | | | | | | | |
|---|---|---|---|---|---|---|---|
| $G_T$ | | $G_E$ | | | $G_R$ | | $G_S$ |
| 0,068 | | 0,445 | | | 0,132 | | 0,355 |
| $g_{T1}$ | $g_{T2}$ | $g_{E1}$ | $g_{E2}$ | $g_{E3}$ | $g_{R1}$ | $g_{R2}$ | $g_{S1}$ |
| $W_{11}$ | $W_{12}$ | $W_{21}$ | $W_{22}$ | $W_{23}$ | $W_{31}$ | $W_{32}$ | $W_{41}$ |
| 0,108 | 0,037 | 0,070 | 0,147 | 0,147 | 0,065 | 0,117 | 0,308 |

| Strengthening | | | | Weakening | Antagonism |
|---|---|---|---|---|---|
| $g_{T1}, g_{T2}$ | $g_{T1}, g_{E3}$ | $g_{E2}, g_{S1}$ | $g_{R1}, g_{R2}$ | $g_{E2}, g_{E3}$ | $g_{E2}, g_{R3}$ |
| $W_{11\_12}$ | $W_{11\_23}$ | $W_{22\_41}$ | $W_{31\_32}$ | $W_{22\_23}$ | $W_{22\_32}$ |
| 0,045 | 0,033 | 0,092 | 0,021 | -0,038 | 0,030 |

6. Conclusions

   In this paper we applied a MCDA approach recently introduced in literature, the robust and hierarchical ELECTRE III method, with a specific improvement of the procedure to elicit weights of criteria.
   The method is the conjunction of four methods that are integrated to give robust recommendations with respect to the problem at hand.
   We simulated the decision-making process by using the preference information provided by two different experts acting as Decision Makers (DMs) and through the steps of the methodology we analyse and discuss the preference ranking of six different requalification projects for an iconic building: the Stock Exchange.
   It is worth mentioning the complexity of some aspects of the architectural field, such as the quantification of purely qualitative aesthetic and spatial aspects in numerical variable. Accordingly, the methodology allows the use of heterogeneous units of measures, which permit to express in the most appropriate way the aesthetic and qualitative aspect typical of the architectural field.
   The uncertainty and imprecision, that can occur in defining the family of criteria is handled and taken into account through the definition of preference and indifference thresholds used by ELECTRE III. Moreover, the preference information expressed by the DMs through the robust SRF-II method led to a large set of compatible vectors of weights and interaction coefficients. All these possibilities are taken into account through the use of the SMAA methodology, which provides robust conclusions on the final rankings of the alternatives by means of frequency in a large number of simulations with different compatible parameter vectors of reference, indifference and incomparability between alternatives not only at comprehensive level but also considering a particular macro-criterion.
   We require to our DMs to express a classification of the criteria before individually and then jointly, in order to provide them three different results based on different preference information. Despite the proposed methodology is able to represent the complexity of a decision-making process through the interactions between the criteria, this aspect could be considered also as a critical point: the information required to the decision makers can be very specific in different areas and according to the fact that the DMs are not required to be competent in all aspects, they could have some trouble in expressing the preferential information. In any case, considering interaction between criteria is a further option that the method supplies and it does not definitely mean that the DM is compelled to use it. The DM can take advantage of the possibility to take into account interaction of criteria if she feels that they are relevant for the decision problem and if she feels to be able to supply reliable preference information. Otherwise, the method can be successfully applied without taking into account any interaction between criteria. Another critical point is the need of many meetings with the DMs, which may be a limit for the applicability of the method (but it must be admitted that it is a common problem with these types of approaches). Of course, one has to accept that if a systematic decision aiding methodology has to be applied, it requires a certain commitment by the DMs, because, otherwise, only inaccurate, misleading and shallow results will be obtained that have to be considered definitely unreliable. On the basis of these remarks, we introduced an innovation in the procedure to elicit weights of criteria. More precisely, we considered the SRF deck of the cards method and we focused our attention on the z value, representing the ratio between the weight of the most important criteria and the weight of the least important criteria. This is an information rather complex and difficult for the DM. The reason of this is the heterogeneity of the z definition with respect to the other information required to the DM, being a number of blank cards between successive levels of cards representing criteria of the same importance. The intuition is that the greater the number of blank cards the greater the difference between weights of criteria in the contiguous levels. With respect to this type of information, the z value requires a quite different logic and this, in general, can generate confusion and perplexity in the DM. In order to handle this problem, we proposed the SRF-II method to replace the z value with an information more homogenous with the other information required in the elicitation procedure. This is the number of blank cards between the card corresponding to least important criterion and the "zero - level". Our case study proved that the DM feels comfortable when required this information that appears clear and understandable, so that the cognitive effort asked is strongly reduced and more solid and safer results are obtained from the elicitation procedure. Consequently, beyond the specific interest for the specific decision problem we considered in this paper, the SRF-II method proves to be, in general, a valuable improvement useful for all its applications.

Finally, we can consider the methodology a valuable tool for aiding the architectural choices within the real-world decision problems, despite the use of architectural criteria remains a challenge that we will continue to investigate.

Acknowledgements

The authors wish to thank Roberta Taramino and Matteo Gianotti for their collaboration as Decision Makers and for the material provided. The authors intend also to thank the students of the course of Master's degree program in Architecture Construction and City at the Politecnico di Torino, coordinated by professor Isabella M. Lami, that conceived the project alternatives (A,B,C,D,E). Salvatore Corrente and Salvatore Greco acknowledge funding by the FIR of the University of Catania BCAEA3 "New developments in Multiple Criteria Decision Aiding (MCDA) and their application to territorial competitiveness". Salvatore Greco has also benefited of the fund Chance of the University of Catania.

Appendix

With respect to the technical macro-criterion, only one partial pre-order (shown in Figure A.1) of the projects at hand is obtained considering the barycenter of the parameters given in Table A.1.

Table A.1. Barycenter of the parameters with respect to the technical macro-criterion

| $G_T$ | | $G_E$ | | | $G_R$ | | $G_S$ |
|---|---|---|---|---|---|---|---|
| $g_{T1}$ | $g_{T2}$ | $g_{E1}$ | $g_{E2}$ | $g_{E3}$ | $g_{R1}$ | $g_{R2}$ | $g_{S1}$ |
| $W_{11}$ | $W_{12}$ | $W_{21}$ | $W_{22}$ | $W_{23}$ | $W_{31}$ | $W_{32}$ | $W_{41}$ |
| 0,108 | 0,03 | 0,070 | 0,147 | 0,147 | 0,067 | 0,115 | 0,308 |

| Strengthening | | | | Weakening | Antagonism |
|---|---|---|---|---|---|
| $g_{T1}, g_{T2}$ | $g_{T1}, g_{E3}$ | $g_{E2}, g_{S1}$ | $g_{R1}, g_{R2}$ | $g_{E2}, g_{E3}$ | $g_{E2}, g_{R3}$ |
| $W_{11\_12}$ | $W_{11\_23}$ | $W_{22\_41}$ | $W_{31\_32}$ | $W_{22\_23}$ | $W_{22\_32}$ |
| 0,111 | 0,148 | 0,277 | 0,070 | -0,033 | 0,043 |

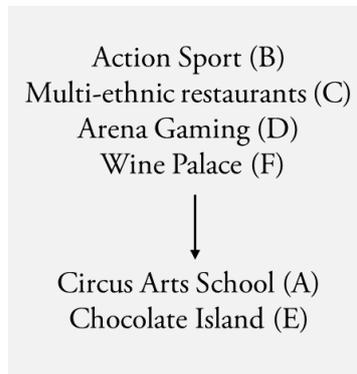

Figure A.1 Unique partial pre-order of the considered projects with respect to the technical macro-criterion

This ranking is supported by the frequencies of preference, indifference and incompatibility shown in Table A.2, Table A.3 and Table A.4, respectively.

Table A.2. Frequency of indifference

|   | A | B | C | D | E | F |
|---|---|---|---|---|---|---|
| A | 0 | 0 | 0 | 0 | 100 | 0 |
| B | 0 | 0 | 100 | 100 | 0 | 100 |
| C | 0 | 100 | 0 | 100 | 0 | 100 |
| D | 0 | 100 | 100 | 0 | 0 | 100 |
| E | 100 | 0 | 0 | 0 | 0 | 0 |
| F | 0 | 100 | 100 | 100 | 0 | 0 |

Table A.3. Frequency of preference

|   | A | B | C | D | E | F |
|---|---|---|---|---|---|---|
| A | 0 | 0 | 0 | 0 | 0 | 0 |
| B | 100 | 0 | 0 | 0 | 100 | 0 |
| C | 100 | 0 | 0 | 0 | 100 | 0 |
| D | 100 | 0 | 0 | 0 | 100 | 0 |
| E | 0 | 0 | 0 | 0 | 0 | 0 |

| | | | | | |
|---|---|---|---|---|---|
| F | 100 | 0 | 0 | 0 | 100 | 0 |

Table A.4. Frequency of incompatibility

| | A | B | C | D | E | F |
|---|---|---|---|---|---|---|
| A | 0 | 0 | 0 | 0 | 0 | 0 |
| B | 0 | 0 | 0 | 0 | 0 | 0 |
| C | 0 | 0 | 0 | 0 | 0 | 0 |
| D | 0 | 0 | 0 | 0 | 0 | 0 |
| E | 0 | 0 | 0 | 0 | 0 | 0 |
| F | 0 | 0 | 0 | 0 | 0 | 0 |

With respect to the economic macro-criterion, three different partial pre-orders (shown in Figures A.2, A.3 and A.4, respectively) can be obtained and the barycenter of the parameters giving these partial pre-orders are shown in Table A.5. According to the barycenter of the weights (Table A.5), the most frequent partial pre-order at secondary level of the hierarchy,

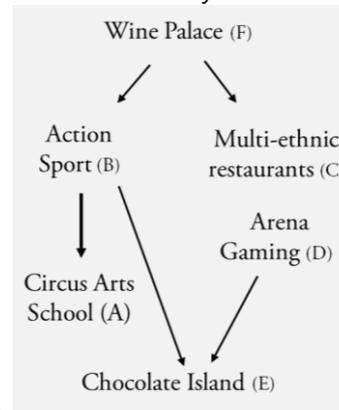

respect to the economic macro-criteria, is shown in
Figure A.2 or occurred in 50,35% of cases, followed by the second partial pre-order with a frequency of 49,38% (Figure A.3) and the third one with a frequency of 10,67%, in Figure A.4.

Table A.5. Barycenters of the parameters on economic macro-criterion

| Most frequent ranking | | | | | | | |
|---|---|---|---|---|---|---|---|
| $G_T$ | | $G_E$ | | | $G_R$ | | $G_S$ |
| $g_{T1}$ | $g_{T2}$ | $g_{E1}$ | $g_{E2}$ | $g_{E3}$ | $g_{R1}$ | $g_{R2}$ | $g_{S1}$ |
| $W_{11}$ | $W_{12}$ | $W_{21}$ | $W_{22}$ | $W_{23}$ | $W_{31}$ | $W_{32}$ | $W_{41}$ |
| 0,109 | 0,038 | 0,072 | 0,145 | 0,145 | 0,067 | 0,116 | 0,308 |

| Strengthening | | | | Weakening | Antagonism |
|---|---|---|---|---|---|
| $g_{T1}, g_{T2}$ | $g_{T1}, g_{E3}$ | $g_{E2}, g_{S1}$ | $g_{R1}, g_{R2}$ | $g_{E2}, g_{E3}$ | $g_{E2}, g_{R3}$ |
| $W_{11\_12}$ | $W_{11\_23}$ | $W_{22\_41}$ | $W_{31\_32}$ | $W_{22\_23}$ | $W_{22\_32}$ |
| 0,062 | 0,113 | 0,227 | 0,049 | -0,035 | 0,050 |

| Second most frequent ranking | | | | | | | |
|---|---|---|---|---|---|---|---|
| $G_T$ | | $G_E$ | | | $G_R$ | | $G_S$ |
| $g_{T1}$ | $g_{T2}$ | $g_{E1}$ | $g_{E2}$ | $g_{E3}$ | $g_{R1}$ | $g_{R2}$ | $g_{S1}$ |
| $W_{11}$ | $W_{12}$ | $W_{21}$ | $W_{22}$ | $W_{23}$ | $W_{31}$ | $W_{32}$ | $W_{41}$ |

| 0,108 | 0,037 | 0,067 | 0,149 | 0,149 | 0,067 | 0,115 | 0,308 |

| Strengthening | | | | Weakening | | Antagonism | |
|---|---|---|---|---|---|---|---|
| $g_{T1}, g_{T2}$ | $g_{T1}, g_{E3}$ | $g_{E2}, g_{S1}$ | $g_{R1}, g_{R2}$ | $g_{E2}, g_{E3}$ | | $g_{E2}, g_{R3}$ | |
| $W_{11\_12}$ | $W_{11\_23}$ | $W_{22\_41}$ | $W_{31\_32}$ | $W_{22\_23}$ | | $W_{22\_32}$ | |
| 0,159 | 0,184 | 0,329 | 0,092 | -0,032 | | 0,037 | |

| Third most frequent ranking | | | | | | | |
|---|---|---|---|---|---|---|---|
| $G_T$ | | $G_E$ | | | $G_R$ | | $G_S$ |
| $g_{T1}$ | $g_{T2}$ | $g_{E1}$ | $g_{E2}$ | $g_{E3}$ | $g_{R1}$ | $g_{R2}$ | $g_{S1}$ |
| $W_{11}$ | $W_{12}$ | $W_{21}$ | $W_{22}$ | $W_{23}$ | $W_{31}$ | $W_{32}$ | $W_{41}$ |
| 0,109 | 0,035 | 0,080 | 0,142 | 0,142 | 0,067 | 0,113 | 0,312 |

| Strengthening | | | | Weakening | | Antagonism | |
|---|---|---|---|---|---|---|---|
| $g_{T1}, g_{T2}$ | $g_{T1}, g_{E3}$ | $g_{E2}, g_{S1}$ | $g_{R1}, g_{R2}$ | $g_{E2}, g_{E3}$ | | $g_{E2}, g_{R3}$ | |
| $W_{11\_12}$ | $W_{11\_23}$ | $W_{22\_41}$ | $W_{31\_32}$ | $W_{22\_23}$ | | $W_{22\_32}$ | |
| 0,014 | 0,055 | 0,164 | 0,045 | -0,049 | | 0,074 | |

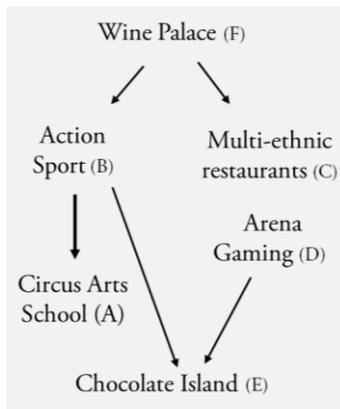

*Figure A.2. Most frequent partial pre-order (50,35%) with respect to the economic macro-criterion*

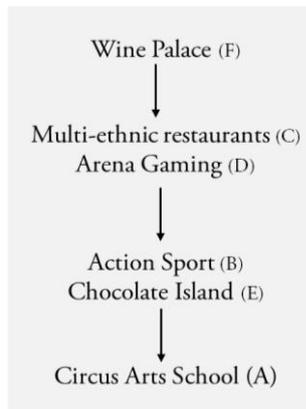

*Figure A.3. Second most frequent partial pre-order (49,38%) with respect to the economic macro-criterion*

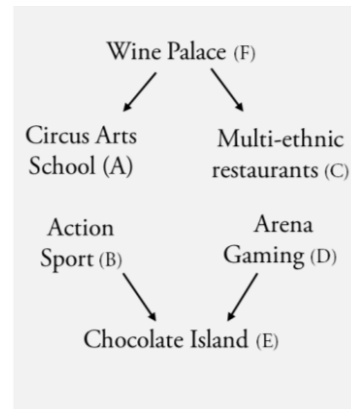

*Figure A.4. Third most frequent partial pre-order (49,38%) with respect to the economic macro-criterion*

The information gathered by these rankings is enriched by the frequencies of preference, indifference and incompatibility shown in Table A.6, A.7, and A.8, respectively.

*Table A.6. Frequency of indifference*

|   | A | B | C | D | E | F |
|---|---|---|---|---|---|---|
| A | 0 | 0,27 | 0 | 0 | 0 | 0 |
| B | 0,27 | 0 | 0 | 0 | 49,38 | 0 |
| C | 0 | 0 | 0 | 100 | 0 | 0 |
| D | 0 | 0 | 100 | 0 | 0 | 0 |
| E | 0 | 49,38 | 0 | 0 | 0 | 0 |
| F | 0 | 0 | 0 | 0 | 0 | 0 |

Table A.7. Frequency of preference

| | A | B | C | D | E | F |
|---|---|---|---|---|---|---|
| A | 0 | 0 | 0 | 0 | 0,27 | 0 |
| B | 99,73 | 0 | 0 | 0 | 50,62 | 0 |
| C | 49,38 | 49,38 | 0 | 0 | 100 | 0 |
| D | 49,38 | 49,38 | 0 | 0 | 100 | 0 |
| E | 49,38 | 0 | 0 | 0 | 0 | 0 |
| F | 100 | 100 | 100 | 100 | 100 | 0 |

Table A.8. Frequency of incompatibility

| | A | B | C | D | E | F |
|---|---|---|---|---|---|---|
| A | 0 | 0 | 50,62 | 50,62 | 50,35 | 0 |
| B | 0 | 0 | 50,62 | 50,62 | 0 | 0 |
| C | 50,62 | 50,62 | 0 | 0 | 0 | 0 |
| D | 50,62 | 50,62 | 0 | 0 | 0 | 0 |
| E | 50,35 | 0 | 0 | 0 | 0 | 0 |
| F | 0 | 0 | 0 | 0 | 0 | 0 |

With respect to the reuse macro-criterion, only one partial pre-order can be obtained (Figure A.5). The barycenter of the parameters giving such a partial pre-order is shown in Table A.9.

Table A.9. Barycenter of the parameters on reuse macro-criterion

| $G_T$ | | $G_E$ | | | $G_R$ | | $G_S$ |
|---|---|---|---|---|---|---|---|
| $g_{T1}$ | $g_{T2}$ | $g_{E1}$ | $g_{E2}$ | $g_{E3}$ | $g_{R1}$ | $g_{R2}$ | $g_{S1}$ |
| $W_{11}$ | $W_{12}$ | $W_{21}$ | $W_{22}$ | $W_{23}$ | $W_{31}$ | $W_{32}$ | $W_{41}$ |
| 0,108 | 0,037 | 0,070 | 0,147 | 0,147 | 0,067 | 0,116 | 0,308 |

| Strengthening | | | | Weakening | Antagonism |
|---|---|---|---|---|---|
| $g_{T1}, g_{T2}$ | $g_{T1}, g_{E3}$ | $g_{E2}, g_{S1}$ | $g_{R1}, g_{R2}$ | $g_{E2}, g_{E3}$ | $g_{E2}, g_{R3}$ |
| $W_{11\_12}$ | $W_{11\_23}$ | $W_{22\_41}$ | $W_{31\_32}$ | $W_{22\_23}$ | $W_{22\_32}$ |
| 0,111 | 0,148 | 0,277 | 0,070 | -0,033 | 0,043 |

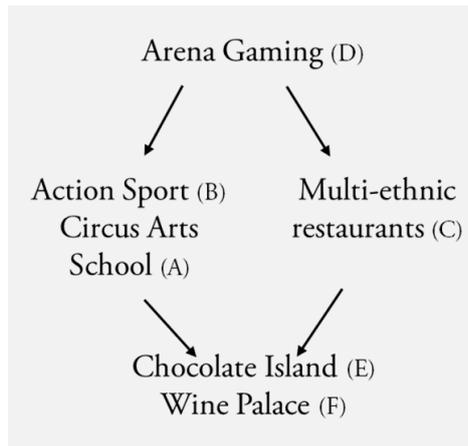

*Figure A.5. Unique partial pre-order with respect to the reuse macro-criterion*

This ranking is supported by the frequencies of preference, indifference and incomparability shown in Tables A.10, A.11 and A.12, respectively.

*Table A. 10. Frequency of indifference*

|   | A | B | C | D | E | F |
|---|---|---|---|---|---|---|
| A | 0 | 100 | 0 | 0 | 0 | 0 |
| B | 100 | 0 | 0 | 0 | 0 | 0 |
| C | 0 | 0 | 0 | 0 | 0 | 0 |
| D | 0 | 0 | 0 | 0 | 0 | 0 |
| E | 0 | 0 | 0 | 0 | 0 | 100 |
| F | 0 | 0 | 0 | 0 | 100 | 0 |

*Table A. 11. Frequency of preference*

|   | A | B | C | D | E | F |
|---|---|---|---|---|---|---|
| A | 0 | 0 | 0 | 0 | 100 | 100 |
| B | 0 | 0 | 0 | 0 | 100 | 100 |
| C | 0 | 0 | 0 | 0 | 100 | 100 |
| D | 100 | 100 | 100 | 0 | 100 | 100 |
| E | 0 | 0 | 0 | 0 | 0 | 0 |
| F | 0 | 0 | 0 | 0 | 0 | 0 |

*Table A. 12. Frequency of incomparability*

|   | A | B | C | D | E | F |
|---|---|---|---|---|---|---|
| A | 0 | 0 | 100 | 0 | 0 | 0 |
| B | 0 | 0 | 100 | 0 | 0 | 0 |
| C | 100 | 100 | 0 | 0 | 0 | 0 |
| D | 0 | 0 | 0 | 0 | 0 | 0 |
| E | 0 | 0 | 0 | 0 | 0 | 0 |
| F | 0 | 0 | 0 | 0 | 0 | 0 |

Also on social macro-criterion a unique partial pre-order of the projects at hand can be obtained (Figure A.6). The barycenter of the parameters giving such a partial pre-order is shown in Table A.13.

*Table A.13. Barycenter of the parameters*

| $G_T$ | | $G_E$ | | | $G_R$ | | $G_S$ |
|---|---|---|---|---|---|---|---|
| $g_{T1}$ | $g_{T2}$ | $g_{E1}$ | $g_{E2}$ | $g_{E3}$ | $g_{R1}$ | $g_{R2}$ | $g_{S1}$ |
| $W_{11}$ | $W_{12}$ | $W_{21}$ | $W_{22}$ | $W_{23}$ | $W_{31}$ | $W_{32}$ | $W_{41}$ |
| 0,108 | 0,037 | 0,070 | 0,147 | 0,147 | 0,067 | 0,116 | 0,308 |

| Strengthening | | | | Weakening | Antagonism |
|---|---|---|---|---|---|
| $g_{T1}, g_{T2}$ | $g_{T1}, g_{E3}$ | $g_{E2}, g_{S1}$ | $g_{R1}, g_{R2}$ | $g_{E2}, g_{E3}$ | $g_{E2}, g_{R3}$ |
| $W_{11\_12}$ | $W_{11\_23}$ | $W_{22\_41}$ | $W_{31\_32}$ | $W_{22\_23}$ | $W_{22\_32}$ |
| 0,111 | 0,148 | 0,277 | 0,070 | -0,033 | 0,043 |

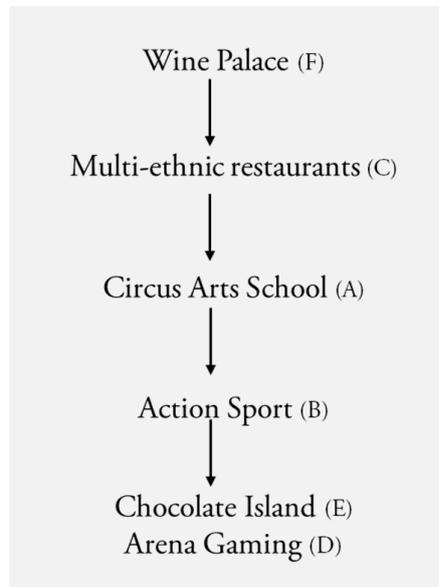

*Figure A.6. Unique partial pre-order with respect to the reuse macro-criterion*

This ranking is supported by the frequencies of preference, indifference and incomparability shown in Tables A.14, A.15 and A.16, respectively.

*Table A.14. Frequency of indifference*

|   | A | B | C | D | E | F |
|---|---|---|---|---|---|---|
| A | 0 | 0 | 0 | 0 | 0 | 0 |
| B | 0 | 0 | 0 | 0 | 0 | 0 |
| C | 0 | 0 | 0 | 0 | 0 | 0 |
| D | 0 | 0 | 0 | 0 | 100 | 0 |
| E | 0 | 0 | 0 | 100 | 0 | 0 |
| F | 0 | 0 | 0 | 0 | 0 | 0 |

*Table A.15. Frequency of preference*

|   | A | B | C | D | E | F |
|---|---|---|---|---|---|---|
| A | 0 | 100 | 0 | 100 | 100 | 0 |
| B | 0 | 0 | 0 | 100 | 100 | 0 |
| C | 100 | 100 | 0 | 100 | 100 | 0 |
| D | 0 | 0 | 0 | 0 | 0 | 0 |
| *E* | 0 | 0 | 0 | 0 | 0 | 0 |
| *F* | 100 | 100 | 100 | 100 | 100 | 0 |

*Table A.16. Frequency of incompatibility*

|   | *A* | *B* | *C* | *D* | *E* | *F* |
|---|---|---|---|---|---|---|
| *A* | 0 | 0 | 0 | 0 | 0 | 0 |
| *B* | 0 | 0 | 0 | 0 | 0 | 0 |
| *C* | 0 | 0 | 0 | 0 | 0 | 0 |
| *D* | 0 | 0 | 0 | 0 | 0 | 0 |
| *E* | 0 | 0 | 0 | 0 | 0 | 0 |
| *F* | 0 | 0 | 0 | 0 | 0 | 0 |

Other data related to the information provided by the two DMs separately together with the obtained partial pre-orders and the frequencies of indifference, preference and incomparability at comprehensive level as well as on the four macro-criteria singularly can be downloaded clicking on the following link: supplementary material.